# Near-field edge fringes at sharp material boundaries

VIKTORIIA E. BABICHEVA*, SAMPATH GAMAGE, MARK I. STOCKMAN, AND YOHANNES ABATE*

*Center for Nano-Optics and Department of Physics and Astronomy, Georgia State University,
P.O. Box 3965, Atlanta 30302, Georgia*
*\* baviev@gmail.com, yabate@gsu.edu*

**Abstract:** We have studied the formation of near-field fringes when sharp edges of materials are imaged using scattering-type scanning near-field optical microscope (s-SNOM). Materials we have investigated include dielectrics, metals, near-perfect conductor, and those that possess anisotropic permittivity and hyperbolic dispersion. For our theoretical analysis, we use a technique that combines full-wave numerical simulations of tip-sample near-field interaction and signal demodulation at higher orders akin to what is done in typical s-SNOM experiments. Unlike previous tip-sample interaction near-field models, our advanced technique allows simulation of the realistic tip and sample structure. Our analysis clarifies edge imaging of recently emerged layered materials such as hexagonal boron nitride and transition metal dichalcogenides (in particular, molybdenum disulfide), as well as traditional plasmonic materials such as gold. Hexagonal boron nitride is studied at several wavelengths, including the wavelength where it possesses excitation of phonon-polaritons and hyperbolic dispersion. Based on our results of s-SNOM imaging in different demodulation orders, we specify resonant and non-resonant types of edges and describe the edge fringes for each case. We clarify near-field edge-fringe formation at material sharp boundaries, both outside bright fringes and the low-contrast region at the edge, and elaborate on the necessity of separating them from propagating waves on the surface of polaritonic materials.

## 1. Introduction

Recently emerged layered materials, such as black phosphorous, transition metal dichalcogenides (TMDCs), and hexagonal boron nitride (hBN), are promising for a wide range of applications in optoelectronics, and their optical properties at the nanoscale have generated enormous interest [1-7]. Hyperbolic metamaterials have recently attracted tremendous attention [8-13] due to potential applications in strong and broadband spontaneous emission and absorption enhancement [14-17], anomalous heat transfer [18], and slow light [19,20] as well as designs of waveguides [21-23] and hyperlenses [24,25]. However, most of these studies are related to artificially designed materials, some of which have to incorporate metal elements that suffer high optical losses. Recently discovered layered materials with natural hyperbolic dispersion [26], such as hBN in the mid-infrared wavelength range [2], open up the possibility of designing devices with better functionality, e.g. less loss of the transmitted signal and higher optical resolution, and consequently may result in real-life applications. Scattering-type scanning near-field optical microscope (s-SNOM) has proved to be an important tool in optical characterizations of such material surfaces providing direct real-space high-resolution images of surface states [27-40].

In s-SNOM experiments, near-field fringe formation at the sample edge is a rather complicated phenomenon for several reasons including but not limited to excitations of edge states and resonances in the structure, and multiple hot spots, as well as image artifacts due to feedback-related issues during rapid scanning around edges [41]. Theoretically, since the edge fringes are formed due to the complex interaction of a curved tip apex and sample, any reliable modeling should include a realistic tip size and shape. Simplified models that approximate the conical probe tip as a point dipole or spheroid [7] provide information about propagating surface waves on sample surfaces, but they cannot reproduce edge fringes that appear at the sample edge. Our work clarifies the origin of edge fringes in polarizable samples and plasmon or phonon resonant materials such as hBN, which supports both bulk and surface hyperbolic waves [2,7,42,43].

Here, we study layered materials with different permittivities and demonstrate an approach to identify material types based on the s-SNOM image of sample edge. We develop a theoretical approach to predict and interpret s-SNOM results at different demodulation orders. When the tip is placed close to a sample surface and illuminated by a laser beam, a complex optical interaction between the tip and the sample takes place. Light is highly concentrated locally at the tip-sample junction (hot spot, Fig. 1), and the scattered field strongly depends on the gap size. The scattered signal depends on tip parameters (geometry and permittivity), sample properties, and the distance between the tip and the sample. We perform full-wave numerical simulations based on the finite-element method and take into account the tip-sample near-field interaction for any shape of the structure, not only flat surface, without any fitting parameters and not being restricted by semi-analytical point-dipole [40] or finite-dipole approximations [37].

We found that metal edges have pronounced bright fringes, whereas edges of dielectrics do not show near-field edge fringes. Similar behavior is observed for layered van der Waals material hBN: bright edge fringes arise in the wavelength range where its dispersion is hyperbolic, and the fringes are absent at frequencies where the material is dielectric. Gold fringes in the visible range are affected by the plasmonic resonances near the edge, and we refer to it as *edge resonances*. These fringes are different from those in the mid-infrared range, where gold is a near-perfect conductor and does not support plasmonic resonances.

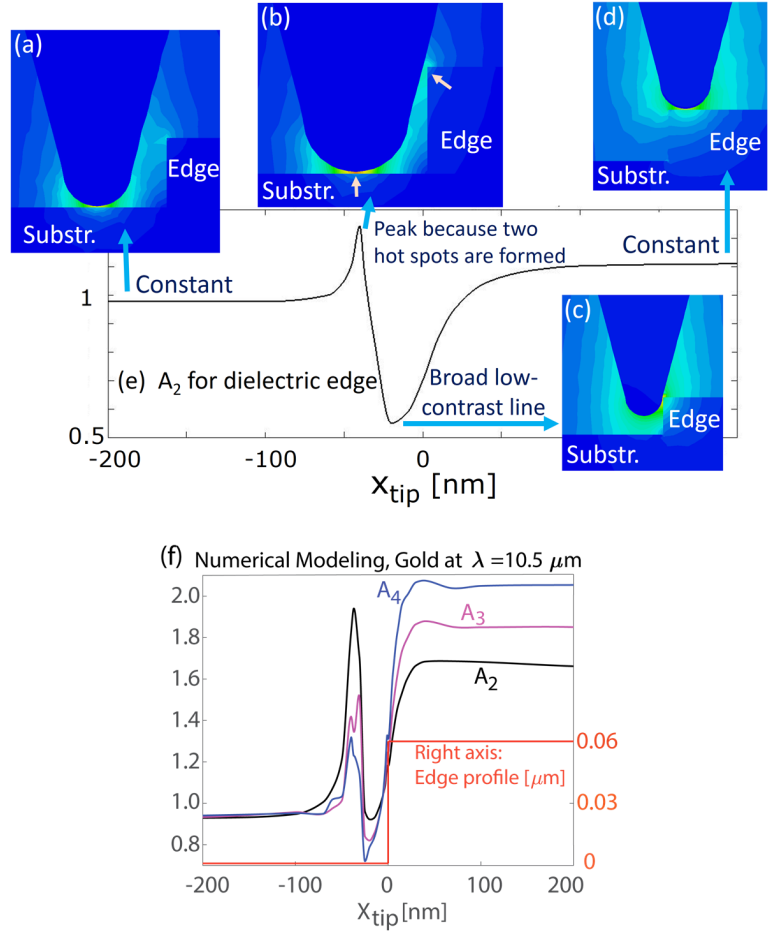

Fig. 1. Numerical calculations showing edge imaging in s-SNOM. (a)-(d) Various tip positions with respect to the dielectric edge and (e) line profile of calculated s-SNOM signal at 2$^{nd}$ harmonic ($A_2$) demodulation. In (b), $x_{tip} = -37$ nm ($x_{tip}$ is a position of the tip with respect to the edge), and two hot spots are formed: below the tip apex and at the edge; this results in the signal peak in (e). In (c) it is shown that the tip is displaced from the substrate below it and consequently does not couple to the substrate strongly, which results in the lower effective polarizability and the broad low-contrast line in the s-SNOM measurements. (f) Change of contrast in different demodulation orders $A_n$ for gold in the mid-infrared range, where it has properties similar to the perfect electric conductor. The signal is normalized so that it is equal to 1 on the substrate at the distance 2 μm in each demodulation order. At the tip position $x_{tip} = -200$ nm, the edge slightly affects the signal, which results in a small deviation from 1.

## 2. Results and discussion

In our theoretical modeling, we consider a realistic conical shape of the tip illuminated by a plane electromagnetic wave. We account for the full structure, including the edge of the sample and the substrate, to realistically simulate the experiment. We then use the height-dependent reflectivity $r^{(CST)}(h)$ calculated by the CST Microwave Studio (frequency-domain solver) to evaluate the effective tip-sample polarizability $\alpha_{\text{eff}}(h) = ia_1 a_2 (r^{(CST)}(h)) * \exp[-2ik_0 \cos\theta(z_{\max} - z_{\text{pos}})] / (2\pi k_0 \tan\theta)$, where $a_1$ and $a_2$ are the dimensions of the simulation box along the x- and y-axes, respectively (periodic boundary conditions are imposed), $k_0$ is the wave vector of the incident beam, $\theta$ is the angle of incidence, $z_{max}$ is the upper boundary of the simulation box, and $z_{pos}$ is the coordinate of the substrate surface. To extract the near-field response of the structure, it is essential to suppress contributions from other kinds of reflection and scattering that do not involve the near-field tip-sample interaction (e.g. reflection from tip shank and substrate). Therefore, the effective polarizability obtained in full-wave simulations is used to calculate the far-field radiation given by $E_s(h_0 + \Delta h \sin \omega_T t) = \sum_n s_n e^{in\omega_T t}$, where $s_n$ is the complex s-SNOM signal, $n$ is the demodulation order, and $\omega_T$ is the natural frequency of the probe tip (typically on the order of 0.3 MHz). To simulate the experiments, we combine $E_S$ with a reference field, $E_M$, that is reflected from an oscillating mirror, implementing a pseudoheterodyne interferometer scheme. To reproduce the experimental measurements, we theoretically demodulate the normalized amplitude s-SNOM signal, which is extracted by dividing the signal at the sample by that at the substrate: $A_n = |s_{n,\text{sample}}|/|s_{n,\text{substrate}}|$.

We perform full-wave simulations of the structure at different tip positions along the sample edge and calculate the signal in various demodulation orders. In combination with interferometric detection similar to the experiments, near-field signal free from background scattering is extracted from higher demodulation orders. At all demodulation orders, we observe a strong field enhancement at small tip-sample distances and a quickly decreasing signal as the tip moves away from the surface reproducing the experiments. Using our model, we analyze different materials and demodulation orders and investigate the formation of near-field edge fringes. In the near-field image of the edge, we specify two main features: a broad line of decreased signal and an outside bright fringe as described in more detail below.

First, we explain the low-contrast line which forms at the edge of a material in s-SNOM images. A cantilevered tip vibrating at a near-resonance frequency is brought in close proximity to a sample surface. As the oscillating cantilever begins to approach the surface, the cantilever oscillation is reduced. The reduction in oscillation amplitude is used to identify and measure surface features [44], and a feedback loop maintains a constant tip-sample separation distance. In the theoretical model, we consider the tip oscillation amplitude of 60-nm and tip-sample distance of 4 nm at any closest point. When the probe tip approaches the edge of the structure, due to the curved profile of the tip, its apex does not couple to the substrate right beneath it (Fig. 1c). The surrounding environment effectively possesses a smaller refractive index, so the effective polarizability of the tip is decreased; thus, a low-contrast line is formed in the s-SNOM image. Such low-contrast line is expected in s-SNOM images of the edge of materials that are sharp and thicker than a few tens of nanometers.

Second, we describe the outside bright fringe in non-resonant materials such as dielectrics and perfect electric conductors. When the tip scans the edge of a sharp sample boundary, there is a point where two hot spots are formed: one at the substrate below the tip and one at the upper sharp edge of the sample, as shown in Fig. 1b. The presence of two hot spots increases the scattered signal; consequently, the bright fringe is formed just outside the geometric edge of the sample. As the tip further moves away from the edge, only one hot spot – between the substrate and bottom of the tip – remains, and in the simulations the s-SNOM signal in the substrate region is nearly independent of the tip position (Fig. 1a). At higher demodulation orders, the bright fringe at the edge of dielectric materials weakens (Fig. 2a,b). For the side hot spot, the tip-scattered field changes slower than for the hot spot formed below the tip.

The sample edge height and the tip shape define the position $x_{tip}$ where two hot spots are simultaneously formed. Exfoliable materials usually have very sharp edges, and the outside fringes are more pronounced in comparison to the edges of other materials with similar permittivity. For thick exfoliable materials (thickness > ~ 50 nm), the edge may have several plateaus, which are clearly seen in the topography line profile (Fig. 2c). The presence of plateaus could result in multiple outside fringes, as shown in the experimental near-field amplitude images of hBN on silicon at $\lambda = 10.7$ μm (Fig. 2c,d) where hBN is an anisotropic dielectric. In the mid-infrared wavelength range, gold produces strong near-field coupling with the probe tip, resulting in large s-SNOM amplitude contrast (Fig. 1f). In this range, optical properties of the gold are similar to the properties of the perfect electric conductor: the real part of the permittivity is negative, while the absolute values of real and imaginary parts of the permittivity are large, so the skin depth is extremely small. Because of this, no plasmonic resonances are involved, the outside bright fringe corresponds to the position of the tip where two hot spots are formed, and the fringe weakens at higher demodulation orders of s-SNOM measurements. Thus, gold edge in the mid-infrared range is non-resonant.

For materials with plasmonic properties or with hyperbolic dispersion, the tip excites edge, and the efficiency of the excitations changes with the tip position. This results in pronounced changes in the scattered near-fields. The origin of the edge resonances can be any resonance in the structure where the resonance is largely defined by the distribution of the fields near the edge. We demonstrate edge resonances for a plasmonic metal (gold in the visible wavelength range) and hBN at mid-IR wavelength, a material with hyperbolic dispersion that supports phonon-polaritons and their multiple reflections within the layer.

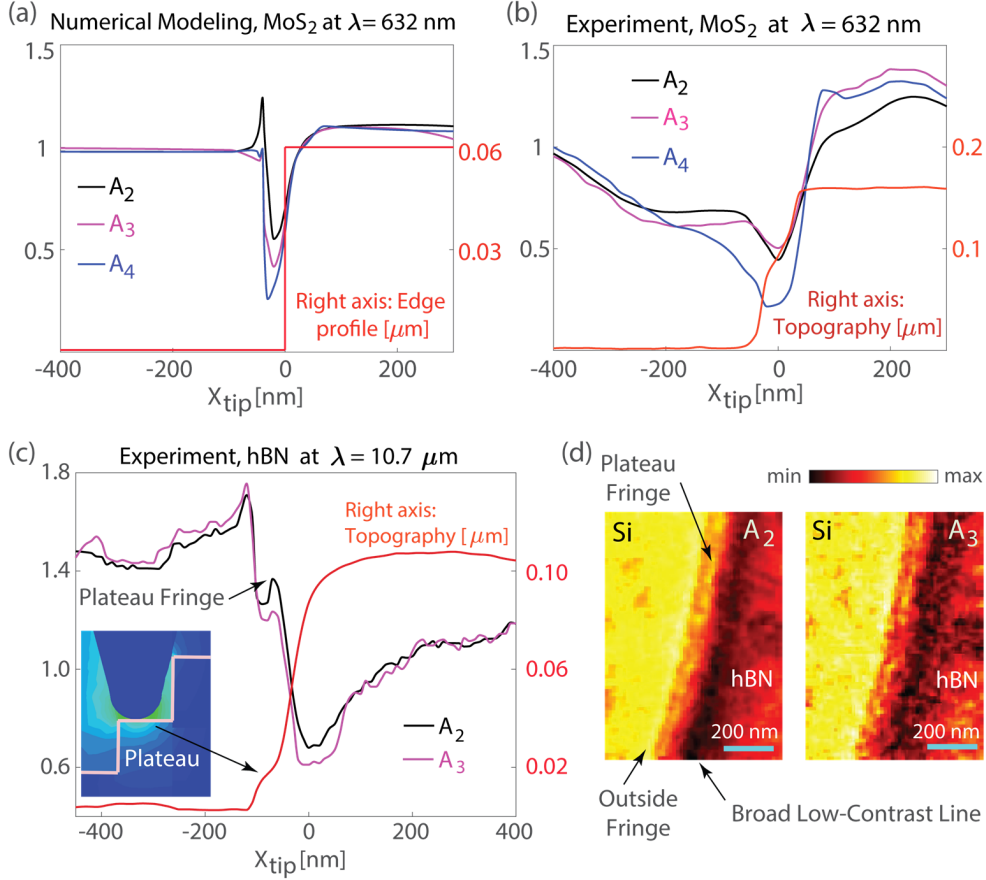

Fig. 2. Edge fringes in materials at different harmonics of the tip resonance frequency $(A_{norm})_n$. (a) and (b) Numerical modeling and experimental images, respectively, for a MoS$_2$ (dielectric) on silicon at λ = 632 nm. (c) Line profiles taken from s-SNOM near-field images and (d) s-SNOM amplitude images for hBN on silicon at laser wavelength, λ = 10.7 μm, where hBN is an anisotropic dielectric. The sample edge has a plateau, which results in an additional outside plateau fringe.

For resonant materials, the brightest point at the edge may appear somewhere in the intermediate position of the tip (as shown in Fig. 1c). This is especially pronounced for hBN with hyperbolic dispersion (Fig. 3c,d). In particular, we consider the excitation wavelength λ = 7 μm, where hBN permittivity components are ($\varepsilon_{in-plane}$ = –14.6 + 1i and $\varepsilon_{out-of-plane}$ = 2.7 + 0.0004i), which results in the hyperbolic dispersion. Out-of-plane component of the permittivity tensor is relatively small, as a result no fringe is observed related to two-hot-spots for hyperbolic hBN. The bright fringes in hBN s-SNOM images are related to the most efficient exaction of hyperbolic phonon-polaritons and resonant reflection of the ray in the layer (Fig. 3a) [42,43]. The direction of the rays is defined by the tensor components of the permittivity (see analytical expressions in [9,42,43]) and thus the strongest excitation occurs only at the particular position of the tip with respect to the edge.

Plasmonic materials, such as gold in the visible wavelength range, have a more complex fringe profile with two peaks (Fig. 3e,f). The peak for negative $x_{tip}$ corresponds to two-hot-spot excitation and strong reflection from them. The second peak results from an efficient excitation of plasmonic edge resonances. This resonance is similar to other plasmonic excitation which occur in nanostructures, where permittivity of metal is comparable with permittivity of surrounding dielectrics [45,46]. However, our theory model does not provide accurate results for this case as the model is limited by the dipole approximation of the tip scattering and does not account for higher multipoles, which can be important in the case of plasmonic nanostructures of complex shape.

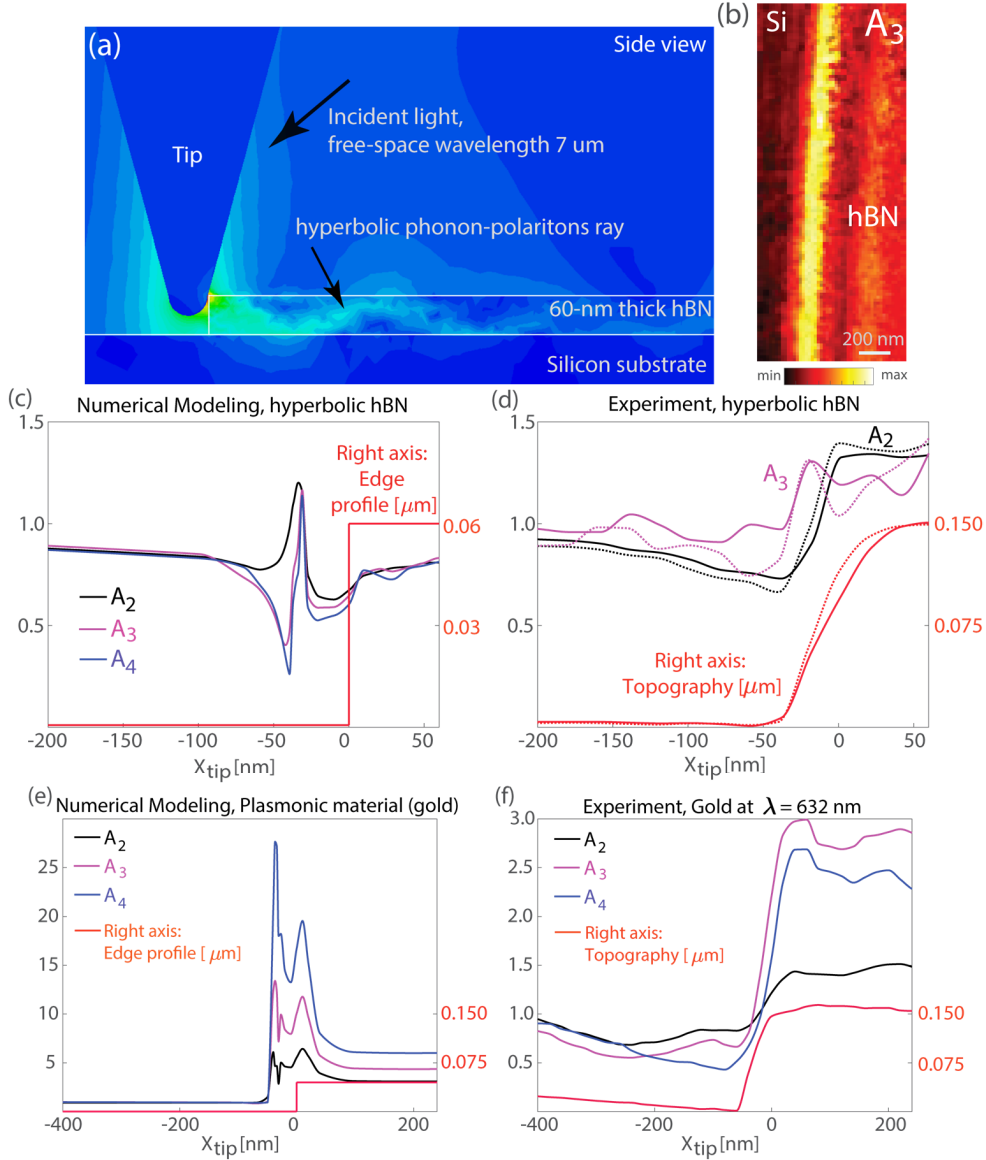

Fig. 3. Edge fringes in plasmonic or phononic structures. (a) Schematic side view of a cross-section of the edge characterization of hBN sample and propagation of deeply subwavelength phonon-polariton rays in 60-nm-thick layer excited by the light with λ = 7 μm. The most efficient excitation and pronounced outside bright fringe occur at the intermediate position of the tip in respect to the sample edge: $x_{tip}$ = –31 nm, which in the numerical simulations corresponds to the peaks in the signal ($A_3$ and $A_4$) of hBN edge characterization. (b) s-SNOM image of hBN at the 3$^{rd}$ harmonic of the tip frequency ($A_3$). hBN fringes in the hyperbolic regime at λ = 7 μm are different from the fringes of dielectric hBN at λ = 10.7 μm (see Fig. 2c,d). (c) Numerical modeling and (d) experimental results of the change of contrast in different demodulation orders $A_n$ for hBN, which has hyperbolic dispersion at λ = 7 μm ($\varepsilon_{in-plane}$ = –14.6 + 1i and $\varepsilon_{out-of-plane}$ = 2.7 + 0.0004i). The peak of the signal occurs at $x_{tip}$ = –31 nm (panel (c)), which means it is offset from the point where two hot spots are created simultaneously (tip/edge and tip/substrate at $x_{tip}$ = –37 nm). In panel (d), the solid and dotted lines correspond to two experimental measurements, and they agree remarkably well. (e) Numerical modeling and (f) experimental results respectively for plasmonic material: at λ = 632 nm, gold permittivity is ε = –12 + 1.3i, and plasmonic edge resonances are excited.

## 3. Conclusion

We have extensively investigated the formation of near-field edge fringes that arise at sharp material edges in s-SNOM measurements. The edge fringes are distinct from other fringes that appear at the sample surface because of the propagation of surface and bulk waves and can give information on the strength of polarizability of materials as well as on their plasmon or phonon resonance edge properties.

## 4. Method

The microscope is a commercial s-SNOM system (neaspec.com). A probing linearly *p*-polarized QCL laser is focused on the tip–sample interface at an angle of $45^0$ to the sample surface. The scattered field is acquired using a phase modulation (pseudoheterodyne) interferometry. The background signal is suppressed by vertical tip oscillations at the mechanical resonance frequency of the cantilever ($f_0 \sim 285$ kHz) and demodulation of the detector signal at higher harmonics $nf_0$, $n = 2, 3, 4$, of the tip resonance frequency. The combined scattered field from the tip and the reference beam pass through a linear polarizer, which further selects the p/p polarization of the measured signal for analysis.

## Funding

SG and YA acknowledge support by the National Science Foundation under grant no. 1553251. The work of VEB is supported by the Air Force Office of Scientific Research (AFOSR) grant number FA9559-16-1-0172. MIS acknowledges support from the Materials Sciences and Engineering Division of the Office of the Basic Energy Sciences, Office of Science, U.S. Department of Energy, Grant No. DE-SC0007043.

## Acknowledgment

The authors acknowledge fruitful discussions with Vladislav Yakovlev.